\documentclass[aps,prl,floatfix,showpacs,twocolumn]{revtex4}
\usepackage{amsmath,amssymb}
\usepackage{epsfig}

\newcommand{\bm}{\boldsymbol}

\begin{document}

\hsize\textwidth\columnwidth\hsize\csname@twocolumnfalse\endcsname

\title{Intrinsic Spin Hall Effect in the presence of Extrinsic Spin-Orbit Scattering}

\author{Wang-Kong Tse} 
\author{S. Das Sarma}
\affiliation{Condensed Matter Theory Center, Department of Physics,
University of Maryland, College Park, Maryland 20742, USA}

\begin{abstract}

Intrinsic and extrinsic spin Hall effects are considered together on an equal theoretical footing for the Rashba spin-orbit coupling in two-dimensional (2D) electron and hole systems, using the diagrammatic method for calculating the spin Hall conductivity. Our analytic theory for the 2D holes shows the expected lowest-order additive result for the spin Hall conductivity. But, the 2D electrons manifest a very surprising result, exhibiting a non-analyticity in the Rashba coupling strength $\alpha$ where the strictly extrinsic spin Hall conductivity (for $\alpha = 0$) cannot be recovered from the $\alpha \to 0$ limit of the combined theory. The theoretical results are discussed in the context of existing experimental results.

\end{abstract}
\pacs{73.43.-f, 72.25.Dc, 75.80.+q, 71.70.Ej}

\maketitle
\newpage

The great deal of current interest in the phenomenon \cite{phys} called the spin Hall effect (SHE) arises from the possibility of controlling spin dynamics in semiconductors using only external electric fields. In particular, the predicted existence of a bulk spin current transverse to the direction of an applied electric field (and hence the charge current) in doped semiconductor structures is both intriguing and interesting. Recent observations of apparent spin accumulation near the edges of doped GaAs 2D and 3D electron \cite{Aws} and 2D hole \cite{Wunder} systems in the presence of an applied electric field have further fuelled this interest as one of the possible explanations for the spin accumulation is that it arises from the SHE-induced bulk spin current although other explanations associated with spin precession effects at the boundaries also exist \cite{bound}. 

Spin Hall effect is traditionally theoretically discussed in terms of two completely distinct physical mechanisms: intrinsic spin Hall effect (ISHE) and extrinsic spin Hall effect (ESHE). ESHE, which is a rather old theoretical prediction, arises from the spin-orbit scattering of semiconductor carriers by impurities (which is known to lead to spatial separation of spin-up and spin-down carriers), and is a solid-state analog of the well-known atomic Mott scattering. ESHE has recently been invoked \cite{ESHE1,ESHEmy} to explain the spin accumulation experiments in the GaAs electron systems although there are still some quantitative discrepancies. The intrinsic effect, ISHE, which is primarily responsible for the current theoretical excitement in the spin Hall effect, is an intrinsic band structure (i.e. periodic lattice) effect, arising entirely from the intrinsic spin-orbit coupling in the band structure of the host semiconductor (i.e. GaAs) material.

Intrinsic and extrinsic spin Hall effects have so far been theoretically discussed completely separately as totally distinct phenomena because of their completely different physical origins: ISHE arising from the intrinsic spin-orbit coupling in the semiconductor band structure and ESHE arising from the extrinsic impurity-induced spin-orbit scattering. This theoretical dichotomy is, however, quite odd since the two mechanisms presumably lead to the same observable effects, namely, a theoretical bulk spin Hall current, and an expreimental boundary spin accumulation. In this Letter, we theoretically study ISHE and ESHE toegther on an equal footing within a single unified SHE formalism. Our work can therefore be considered to be either `intrinsic spin Hall effect in the presence of extrinsic spin-orbit scattering' (as the title of our paper suggests), or equivalently, `extrinsic spin Hall effect in the presence of intrinsic Rashba spin-orbit coupling' (which the title of our paper could easily have been). We consider the well-known and extensively studied Rashba model for the intrinsic spin-orbit coupling in the semiconductor carriers, restricting ourselves to 2D GaAs electron \cite{Aws} and hole \cite{Wunder} systems where the Rashba coupling is expected to be the main band structure spin-orbit coupling mechanism arising from the spatially imposed structural inversion asymmetry in the GaAs system. Our work is apparently the only work in the literature to theoretically treat both ISHE and ESHE on an equal footing within a unified theoretical formalism although there have been many studies of ISHE \cite{phys,ISHE1} and ESHE \cite{ESHE1,ESHEmy} separately.

Our theory treating ISHE and ESHE together uses a minimal model with a parabolic carrier band dispersion, characterised by an effective mass $m$, with four other independent parameters, the Rashba spin-orbit coupling strength $\alpha$ (or equivalently the Rashba spin splitting $\Delta$), the effective extrinsic spin-orbit impurity scattering strength $\lambda_0$, the carrier Fermi energy $\varepsilon_F$ (or equivalently the carrier density $n$), and a charge transport relaxation time $\tau$ (related to the ordinary Drude charge conductivity through $\sigma = ne^2\tau/m$) completely defining the one-electron transport problem. We take into account both the side-jump (SJ) and the skew scattering (SS) contributions to the spin Hall conductivity precisely in a well-defined diagrammatic expansion. Below we describe our theory and results for 2D electrons and holes separately.


\textit{2D electron gas with Rashba interaction.}---We consider the single-particle Hamiltonian in the presence of both Rashba SO coupling and SO scattering due to impurities, the disorder-free part of which is given by:
\begin{eqnarray}
H_0 = \frac{p^2}{2m}+\alpha p\,(\mathrm{sin}\phi\,\sigma_x-\mathrm{cos}\phi\,\sigma_y),
\label{eq1}
\end{eqnarray}
where $\phi = \mathrm{tan}^{-1}(p_y/p_x)$, and the disorder-dependent part which includes the SO scattering is given by: 
\begin{eqnarray}
V_{\mathrm{imp}} = -\frac{\lambda_0^2}{4}
\left[\boldsymbol{\sigma}\times\boldsymbol{\nabla}V(\boldsymbol{r})\right]\cdot\boldsymbol{p}+V(\boldsymbol{r}),
\label{eq2}
\end{eqnarray}
As shown in \cite{ESHEmy}, the effect of SO extrinsic scattering can be taken into account 
by calculating the diagrams with corrections of the spin and charge current vertices (side-jump) and the SO scattering amplitude (skew scattering). 
The second-quantized Hamiltonian reads
\begin{eqnarray}
&&H = \label{eq6} \\
&&\sum_{\bm{k}\bm{k}'} \psi^{\dag}_{\bm{k}}\bigg\{H_0\delta_{\bm{k}\bm{k}'}
+V_{\boldsymbol{k}-\boldsymbol{k}'}\bigg[1-\frac{i\lambda_0^2}{4}(\bm{k}\times\bm{k}')\cdot\bm{\sigma})\bigg]\bigg\}\psi_{\bm{k}'},
\nonumber
\end{eqnarray}
The spin current and charge current read 
\begin{eqnarray}
\boldsymbol{J}_s &=& e\sum_{\bm{k}\bm{k}'}\psi^{\dag}_{\bm{k}}\bigg\{\frac{\hbar \bm{k}}{2m} \sigma_z
\delta_{\bm{k}\bm{k}'} \nonumber \\
&&-\frac{i\lambda_0^2}{8}V_{\boldsymbol{k}-\boldsymbol{k}'}
\left[\bm{\hat{z}}\times\left(\bm{k}-\bm{k}'\right)\right]\bigg\}
\psi_{\bm{k}'}, \label{eq7} \\
\boldsymbol{J}_c &=& e\sum_{\bm{k}\bm{k}'}\psi^{\dag}_{\bm{k}}\bigg\{\frac{\hbar \boldsymbol{k}}{m}\delta_{\bm{k}\bm{k}'}+\alpha(-\sigma_y\hat{\bm{x}}+\sigma_x\hat{\bm{y}}) \nonumber \\
&&-\frac{i\lambda_0^2}{4}V_{\boldsymbol{k}-\boldsymbol{k}'}
\bm{\sigma}\times\left(\bm{k}-\bm{k}'\right)\bigg\}
\psi_{\bm{k}'}. \label{eq8} 
\end{eqnarray}
We proceed to calculate the ISHE conductivity $\sigma_{yx}^{SH}$ using the Kubo-Greenwood formula, in the presence of extrinsic spin Hall effect. The retarded and advanced Green functions are given by 
\begin{eqnarray}
G^{(R,A)}_k &=& \frac{1}{(\varepsilon-\xi_k\pm i\hbar/4\tau)^2-(\alpha p)^2} \nonumber \\
&&\left[
\begin{array}{cc}
\varepsilon-\xi_k \pm i\hbar/4\tau & i\alpha p \mathrm{e}^{-i\phi} \\
-i\alpha p \mathrm{e}^{i\phi} & \varepsilon-\xi_k \pm i\hbar/4\tau
\end{array}
\right], \label{eq2}
\end{eqnarray}
where $\xi_k = p^2/2m-\varepsilon_F$. In the dilute impurity limit $\varepsilon_F\tau \gg 1$, diagrams with intersection of the impurity lines can be neglected. We also consider the energy splitting between the two chiral branches $\Delta/2 = 2\alpha p_F \ll \varepsilon_F$ and neglect corrections of the order of $O(\Delta/\varepsilon_F)$. (These are all standard approximations in the context of SHE.)

The diagrams in Fig.~1A, B, E, F and Fig.~2I, J are, respectively, the diagrams for the side-jump and skew scattering contributions. Note that the Green function lines include the Rashba term in the Hamiltonian, Eq.~(\ref{eq2}), i.e. the ESHE is now modified by ISHE. First we calculate the diagrams for the side-jump without diffuson pole vertex correction (Fig.~1A, B, E, F), giving:
\begin{eqnarray}
&&\sigma_{yx}^{A+B} = -\frac{ie^2\lambda_0^2\hbar}{16\pi m}\; n_i v_0^2
\mathrm{tr}\sum_{\bm{k}_1\bm{k}_2} k_{1x}^2G^R_{\bm{k}_1}G^A_{\bm{k}_1}
\left(G^R_{\bm{k}_2}-G^A_{\bm{k}_2}\right) \nonumber \\
&&+\frac{ie^2\lambda_0^2\alpha}{16\pi}\; n_i v_0^2
\mathrm{tr}\sum_{\bm{k}_1\bm{k}_2} k_{1x}G^R_{\bm{k}_1}\sigma_y G^A_{\bm{k}_1}
\left(G^R_{\bm{k}_2}-G^A_{\bm{k}_2}\right).
\label{eq9}
\end{eqnarray}
\begin{eqnarray}
&&\sigma_{yx}^{E+F} = -\frac{ie^2\lambda_0^2\hbar}{16\pi m}\; n_i v_0^2 \nonumber \\
&&\mathrm{tr}\sum_{\bm{k}_1\bm{k}_2} k_{1y}^2 \sigma_z G^R_{\bm{k}_1}
\left(G^R_{\bm{k}_2}\sigma_z-\sigma_z G^A_{\bm{k}_2}\right)G^A_{\bm{k}_1}.
\label{eq10}
\end{eqnarray}
Diagrams in Fig.~1A, B and Fig.~1E, F correspond to the contributions from the vertex renormalizations of the spin current and charge current, respectively, due to the anomalous SO current vertex. In the following we also take into account of vertex corrections due to diffuson poles. We have to consider the vertex corrections to two types of vertices, one for the charge current vertex on the right and the other for the spin current vertex on the left. For the charge current vertex, it is well known that the diffuson pole vertex correction leads to an exact cancellation of the spin-dependent term of the charge current $-\alpha \sigma_y$ for the Rashba model (this is the now well-accepted precise vanishing of the pure ISHE in the 2D Rashba model). Therefore, with diffuson pole vertex correction, the second term on the right side of Eq.~(\ref{eq9}), which corresponds to a current-current correlation of the anomalous SO current and the Rashba spin-dependent current $-\alpha \sigma_y$, is exactly cancelled by an additional term opposite in sign, thus: 
\begin{eqnarray}
&&\sigma_{yx}^{A+B+C+D} \nonumber \\
&=&-\frac{ie^2\lambda_0^2\hbar}{16\pi m}\; n_i v_0^2
\mathrm{tr}\sum_{\bm{k}_1\bm{k}_2} k_{1x}^2G^R_{\bm{k}_1}G^A_{\bm{k}_1}
\left(G^R_{\bm{k}_2}-G^A_{\bm{k}_2}\right).
\label{eq11}
\end{eqnarray}
For the spin current vertex, the vertex correction $\Gamma_{sy}$ satisfies the Bethe-Salpeter equation
\begin{equation}
\Gamma_{sy} = \frac{1}{2\pi\tau N}\sum_{\bm{k}} G^A_{\bm{k}}(\frac{\hbar k_y}{2m}\sigma_z+\Gamma_{sy})G^R_{\bm{k}}.
\label{eq12}
\end{equation}
Here $N$ is the 2D density of states. Evaluating the first term under the integral on the right-hand side yields a term proportional to $\sigma_y$, which suggests that the ansatz $\Gamma_{sy} = \gamma \sigma_y$, with $\gamma$ just a constant to be determined, is a solution of Eq.~(\ref{eq12}). Straightforward calculation gives $\gamma = v_F/2\Delta\tau$, where $v_F$ is the Fermi velocity. Using this spin current vertex correction $v_F\sigma_y/2\Delta\tau$ (Fig.~1G and H), we find the vertex correction to diagrams Fig.~1E and F to be
\begin{eqnarray}
&&\sigma_{yx}^{G+H} = -\frac{ie^2\lambda_0^2v_F}{16\pi \Delta\tau}\; n_i v_0^2 \nonumber \\
&&\mathrm{tr}\sum_{\bm{k}_1\bm{k}_2} k_{1y}\sigma_y G^R_{\bm{k}_1}
\left(G^R_{\bm{k}_2}\sigma_z-\sigma_z G^A_{\bm{k}_2}\right)G^A_{\bm{k}_1}.
\label{eq13}
\end{eqnarray}
\begin{figure}
  \includegraphics[width=8.5cm,angle=0]{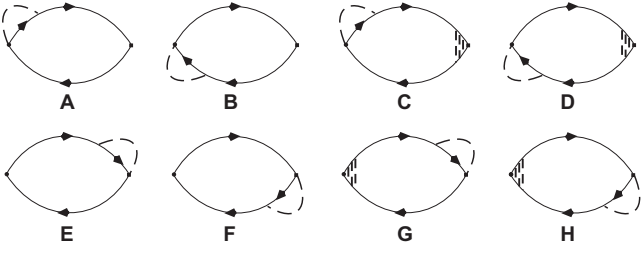}
\caption{Diagrams for the side-jump contribution, where the left vertex 
is for y-component of the spin current and the right vertex is for the x-component of the 
charge current. The vertex
correction due to the anomalous SO vertex is denoted by a dashed line 
connected to that vertex; while the vertex correction due to diffuson pole is by ladder impurity lines.
} \label{fig1}
\end{figure}
Now we proceed to calculate the diagrams for the skew scattering. Without diffuson pole vertex correction, the diagrams Fig.~2I and J give
\begin{eqnarray}
&&\sigma_{yx}^{I+J} = \frac{ie^2\lambda_0^2\hbar^2}{16\pi m^2}\; n_i v_0^3
\mathrm{tr}\sum_{\bm{k}_1\bm{k}_2\bm{k}_3} k_{1y}^2 k_{2x}^2 \sigma_z \label{eq14} \\
&&G^R_{\bm{k}_1}\left(G^R_{\bm{k}_3}G^R_{\bm{k}_2}G^A_{\bm{k}_2}\sigma_z
-\sigma_z G^R_{\bm{k}_2}G^A_{\bm{k}_2}G^A_{\bm{k}_3}\right)
G^A_{\bm{k}_1} \nonumber \\
&&+\frac{ie^2\lambda_0^2\hbar\alpha}{16\pi m}\; n_i v_0^3
\mathrm{tr}\sum_{\bm{k}_1\bm{k}_2\bm{k}_3} k_{1x} k_{1y} k_{2y} \sigma_z \nonumber \\
&&G^R_{\bm{k}_1}\left(G^R_{\bm{k}_3}G^R_{\bm{k}_2}\sigma_y G^A_{\bm{k}_2}\sigma_z
-\sigma_z G^R_{\bm{k}_2}\sigma_y G^A_{\bm{k}_2}G^A_{\bm{k}_3}\right)
G^A_{\bm{k}_1}. \nonumber
\end{eqnarray}
\begin{figure}
  \includegraphics[width=8.5cm,angle=0]{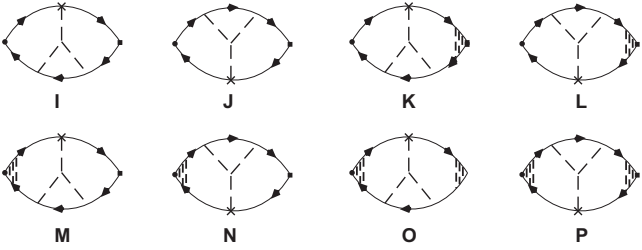}
\caption{Diagrams for the skew-scattering contribution, the cross
denotes a correction to the Green's function Eq.~(\ref{eq2}) due to
SO scattering.} \label{fig2}
\end{figure}
The first sum in Eq.~(\ref{eq14}) corresponds to the current-current correlation of the spin current $\hbar k_{y}/2m$ with the usual charge current $\hbar k_{x}/m$ while the second sum corresponds to that with the Rashba spin-dependent current $-\alpha\sigma_y$. Similar to Eq.~(\ref{eq11}), taking account of the vertex correction to the right-side charge current cancels the second term, so that $\sigma_{yx}^{I+J+K+L}$ is given by only the first term in Eq.~(\ref{eq14}). 

The diagrams for the vertex correction to the spin current Fig.~2M, N give
\begin{eqnarray}
&&\sigma_{yx}^{M+N} = \frac{ie^2\lambda_0^2\hbar v_F}{16\pi m \Delta\tau}\; n_i v_0^3
\mathrm{tr}\sum_{\bm{k}_1\bm{k}_2\bm{k}_3} k_{1y} k_{2x}^2 \sigma_y   \label{eq15} \\
&&G^R_{\bm{k}_1}\left(G^R_{\bm{k}_3}G^R_{\bm{k}_2}G^A_{\bm{k}_2}\sigma_z
-\sigma_z G^R_{\bm{k}_2}G^A_{\bm{k}_2}G^A_{\bm{k}_3}\right)
G^A_{\bm{k}_1} \nonumber \\
&&+\frac{ie^2\lambda_0^2 v_F \alpha}{16\pi \Delta\tau}\; n_i v_0^3
\mathrm{tr}\sum_{\bm{k}_1\bm{k}_2\bm{k}_3} k_{1x} k_{2y} \sigma_y \nonumber \\
&&G^R_{\bm{k}_1}
\left(G^R_{\bm{k}_3}G^R_{\bm{k}_2}\sigma_y G^A_{\bm{k}_2}\sigma_z
-\sigma_z G^R_{\bm{k}_2}\sigma_y G^A_{\bm{k}_2}G^A_{\bm{k}_3}\right)
G^A_{\bm{k}_1}. \nonumber
\end{eqnarray}
Finally, we have the diagrams in Fig.~2O and P which take into account the diffuson pole vertex corrections to both vertices, however, they are also the charge current vertex correction to the diagrams in Fig. 2M and N, which therefore cancels the second term in Eq.~(\ref{eq15}), and $\sigma_{yx}^{M+N+O+P}$ is again given only the first term in Eq.~(\ref{eq15}).

Now we proceed to evaluate the integrals in Eqs.~(\ref{eq10}), (\ref{eq11}), (\ref{eq13}) for side-jump and the first terms on the right-hand side of Eqs.~(\ref{eq14})-(\ref{eq15}) corresponding to skew scattering. Eqs.~(\ref{eq11}) and (\ref{eq13}) give
\begin{equation}
\sigma_{yx}^{E+F} = -\sigma_{yx}^{G+H} = -\frac{me^2\lambda_0^2k_F^2\tau}{8\pi \hbar^4} n_i v_0^2 
\frac{1}{1+(\Delta\tau)^2},
\label{eq16}
\end{equation}
while Eqs.~(\ref{eq14})-(\ref{eq15}) give
\begin{eqnarray}
\sigma_{yx}^{I+J+K+L} &=& -\sigma_{yx}^{M+N+O+P} \nonumber \\
&=& \frac{me^2\lambda_0^2k_F^4\tau^2}{8\pi \hbar^6}n_i v_0^3
\frac{1}{1+(\Delta\tau)^2}.
\label{eq17}
\end{eqnarray}
We note that these terms correspond to a combined action of the intrinsic and extrinsic spin Hall effect, as manifested in the appearance of both $\Delta$ and $n_i$. However, somewhat unexpectedly, the spin current vertex corrections Fig.~1G and H cancel the side-jump contributions Fig.~1E and F. More importantly, we find that the entire contribution to the skew scattering vanishes in the presence of vertex corrections as well. Now we evaluate the remaining Eq.~(\ref{eq11}). This gives nothing but the usual side-jump contribution in the purely extrinsic spin Hall effect:
\begin{equation}
\sigma_{xy}^{SH} = -\sigma_{yx}^{SH} = \sigma_{yx}^{A+B+C+D} = \frac{e^2\lambda_0^2}{8\hbar}n.
\label{eq18}
\end{equation}
We note here that the value is halved as compared to the case of pure ESHE \cite{ESHEmy} since half of the side-jump contribution is cancelled by vertex correction. 

Without taking account of extrinsic spin Hall effect, the intrinsic spin Hall conductivity is $(e^2/8\pi\hbar)(\Delta\tau)^2/[1+(\Delta\tau)^2]$, which is exactly cancelled by the diffuson vertex correction \cite{ISHE1}. 
Now we have shown that, when the spin-orbit scattering from impurities is taken into account in the intrinsic spin Hall effect, there is no residual intrinsic spin Hall term, nor is there any mixed-term correponding to the combined action of the intrinsic and extrinsic spin Hall effects. Moreover, the contribution to the intrinsic spin Hall conductivity from the skew scattering vanishes exactly, and only the side-jump contribution survives. Therefore, in the presence of extrinsic SO scattering, the total (intrinsic $+$ extrinsic) spin Hall conductivity is dominated by only the extrinsic contribution of side-jump: $\sigma_{xy}^{SJ} = \sigma_{xy}^{SJE}/2$ and $\sigma_{xy}^{SS} = 0$, where $\sigma_{xy}^{SJE}$ is the side-jump result of the pure ESHE \cite{ESHEmy}. This result is striking -- it implies that the $\alpha \to 0$ limit is non-analytic, i.e. SHE ($\alpha = 0$) $\equiv$ ESHE $\not=$ SHE ($\alpha \to 0$) = $[\textrm{ISHE} + \textrm{ESHE}]_{\alpha \to 0}$. This non-analyticity, existing only within the Rashba model for 2D electrons, arises from the curious fact that as $\alpha \to 0$, the vertex correction $\Gamma_{sy}$, defined by Eq.~(\ref{eq12}), is non-zero (in fact it diverges), whereas for $\alpha \equiv 0$, $\Gamma_{sy} = 0$ by definition. This non-analyticity shows that the 2D electronic Rashba model is rather special. We also note that the $\lambda_0 \to 0$ limit of the theory is analytic. 


\textit{2D hole gas with Rashba interaction.}---We now consider a two-dimensional gas of heavy holes (with spin $3/2$) with Rashba interaction. In this case the disorder-free part of the Hamiltonian is given by
\begin{equation}
H_0 = \frac{p^2}{2m}+\alpha p^3 \left(\mathrm{sin}3\phi\,\sigma_x-\mathrm{cos}3\phi\,\sigma_y\right). 
\label{eq19}
\end{equation}
The single-particle spin current and charge current operators are
\begin{eqnarray}
j_{sy} &=& e\frac{3\hbar}{2m}k_y\sigma_z, \label{eq20} \\
j_{cx} &=& e\left[\frac{\hbar k_x}{m}+3\alpha k^2\left(\mathrm{sin}2\phi\,\sigma_x-\mathrm{cos}2\phi\,\sigma_y\right)\right], \label{eq21}
\end{eqnarray}
and the Green function in this case is just given by Eq.~(\ref{eq2}) with $\alpha p \to \alpha p^3$ and $\phi \to 3\phi$. First we evaluate the bubble diagram without extrinsic spin-orbit interaction (i.e. purely intrinsic Rashba contribution), which is found to be $\sigma_{xy} = (9e^2/8\pi\hbar)(\Delta\tau)^2/[1+(\Delta\tau)^2]$ (here for holes $\Delta/2 = 2\alpha \hbar^2 k_F^3$). The ladder correction to the charge current vertex is found to be zero. This agrees with earlier results \cite{ISHE2} in the clean limit. The ladder correction to the spin current vertex is given by Eq.~(\ref{eq12}) except with a extra factor of $3$ attached to the bare spin current term. Evaluating the first term on the right hand side of Eq.~(\ref{eq12}) gives zero, because here we encounter terms with $\mathrm{cos}3\phi, \mathrm{sin}3\phi$ instead of  $\mathrm{cos}\phi, \mathrm{sin}\phi$ in the electron case, and when integrated with $\int \mathrm{d}\phi\,\mathrm{sin}\phi$ these terms yield zero by orthogonality. Therefore the ladder correction to the spin current vertex is zero as well, which is an expected result since the diffuson ladder correction diagram can be regarded as renormalization of either the charge current vertex on the right or the spin current vertex on the left. 

Now that all the ladder vertex correction vanishes, we only have the remaining diagrams Fig.~1A, B, E, F and Fig.~2I, J. For side-jump, diagrams A and B yield
\begin{eqnarray}
&&\sigma_{yx}^{A+B} = -\frac{ie^2\lambda_0^2\hbar}{16\pi m}\; n_i v_0^2
\mathrm{tr}\sum_{\bm{k}_1\bm{k}_2} k_{1x}^2G^R_{\bm{k}_1}G^A_{\bm{k}_1}
\left(G^R_{\bm{k}_2}-G^A_{\bm{k}_2}\right) \nonumber \\
&&+\frac{i3e^2\lambda_0^2\alpha}{16\pi}\; n_i v_0^2
\mathrm{tr}\sum_{\bm{k}_1\bm{k}_2} \nonumber \\
&&k_1^2 k_{1x}G^R_{\bm{k}_1}(\mathrm{cos}2\phi\,\sigma_y-\mathrm{sin}2\phi\,\sigma_x) G^A_{\bm{k}_1}
\left(G^R_{\bm{k}_2}-G^A_{\bm{k}_2}\right). \label{eq22}
\end{eqnarray}
while diagrams E and F give
\begin{eqnarray}
&&\sigma_{yx}^{E+F} = -\frac{i3e^2\lambda_0^2\hbar}{16\pi m}\; n_i v_0^2 \nonumber \\
&&\mathrm{tr}\sum_{\bm{k}_1\bm{k}_2} k_{1y}^2 \sigma_z G^R_{\bm{k}_1}
\left(G^R_{\bm{k}_2}\sigma_z-\sigma_z G^A_{\bm{k}_2}\right)G^A_{\bm{k}_1}.
\label{eq23}
\end{eqnarray}
Evaluating Fig.~2I and J for skew scattering gives
\begin{eqnarray}
&&\sigma_{yx}^{I+J} = \frac{i3e^2\lambda_0^2\hbar^2}{16\pi m^2}\; n_i v_0^3
\mathrm{tr}\sum_{\bm{k}_1\bm{k}_2\bm{k}_3} k_{1y}^2 k_{2x}^2 \sigma_z \label{eq24} \\
&&G^R_{\bm{k}_1}\left(G^R_{\bm{k}_3}G^R_{\bm{k}_2}G^A_{\bm{k}_2}\sigma_z
-\sigma_z G^R_{\bm{k}_2}G^A_{\bm{k}_2}G^A_{\bm{k}_3}\right)
G^A_{\bm{k}_1} \nonumber \\
&&+\frac{i9e^2\lambda_0^2\hbar\alpha}{16\pi m}\; n_i v_0^3
\mathrm{tr}\sum_{\bm{k}_1\bm{k}_2\bm{k}_3} k_2^2 k_{1y}^2 k_{2x} \sigma_z \nonumber \\
&&G^R_{\bm{k}_1}\left[G^R_{\bm{k}_3}G^R_{\bm{k}_2}(\mathrm{sin}2\phi_2\,\sigma_x-\mathrm{cos}2\phi_2\,\sigma_y) G^A_{\bm{k}_2}\sigma_z \right. \nonumber \\
&&\left.-\sigma_z G^R_{\bm{k}_2}(\mathrm{sin}2\phi_2\,\sigma_x-\mathrm{cos}2\phi_2\,\sigma_y) G^A_{\bm{k}_2}G^A_{\bm{k}_3}\right)
G^A_{\bm{k}_1}. \nonumber
\end{eqnarray}
Evaluating the integrals, we find the second term in Eq.~(\ref{eq22}) and also in Eq.~(\ref{eq24}) are zero. The total side-jump contribution therefore gives:
\begin{equation}
\sigma_{xy}^{SJ} = -\sigma_{yx}^{SJ} = \frac{e^2\lambda_0^2}{8\hbar}n\left[1+\frac{3}{1+(\Delta\tau)^2}\right].
\label{eq25}
\end{equation}
and the total skew scattering contribution, within the short-range screened impurity assumption \cite{ESHEmy}, gives
\begin{equation}
\sigma_{xy}^{SS} = -\sigma_{yx}^{SS} = -\frac{3\pi m\lambda_0^2 \varepsilon_F}{2\hbar^2}\frac{1}{1+(\Delta\tau)^2}\left(\frac{ne^2\tau}{m}\right)
\label{eq26}
\end{equation}
It can be recognized that the combined effect of both the extrinsic and intrinsic spin Hall effect on heavy holes is to modify the extrinsic spin Hall results by a factor coming from the intrinsic effect. In particular, when $\Delta \to 0$, Eqs.~(\ref{eq25})-(\ref{eq26}) reduce to the results for extrinsic spin Hall effect for heavy holes. The total SHE for holes is therefore additive (i.e. ISHE + ESHE) in the leading order, which is very different from the singular result for 2D electrons.

In summary, we emphasize that in the case of Rashba SO coupling for electrons, the spin Hall conductivity is singular at zero Rashba strength $\alpha = 0$, where there will be a non-zero contribution coming from skew scattering \cite{ESHEmy}. This can be traced back to the singular nature of the ladder vertex correction to the spin current: the solution of Eq.~(\ref{eq12}) is zero for $\alpha \equiv 0$ and yet becomes non-zero with a value of $v_F\sigma_y/2\Delta\tau$ even for an infinitesimal value of $\alpha$. In the case for heavy holes, the spin Hall conductivity is analytic with respect to the Rashba coupling strength and produces the expected perturbative result that in the leading order the net spin Hall effect is a sum of intrinsic and extrinsic SHE in the absence of each other. Finally, the implications of our theory for the experimental results are: (1) for 2D electrons, the ESHE calculated before \cite{ESHEmy} will be enhanced since the skew scattering contribution (with an opposite sign) vanishes, bringing theory and experiment in better agreement; (2) for 2D holes, the additive perturbative result means that both ISHE and ESHE contribute to experiment, with the ISHE being quantitatively larger.

This work is supported by US-ONR and NSF.


\end{document}